\documentclass[twocolumn,pra,aps,superscriptaddress,flushbottom,showpacs]{revtex4-1}
\usepackage{xspace,amsmath,amsfonts,amsthm,amssymb,amsbsy,graphicx,color}

\begin{document} 

\newcommand{\ms}[1]{\mbox{\scriptsize #1}}
\newcommand{\msb}[1]{\mbox{\scriptsize $\mathbf{#1}$}}
\newcommand{\msi}[1]{\mbox{\scriptsize\textit{#1}}}
\newcommand{\nn}{\nonumber} 
\newcommand{\dg}{^\dagger} 
\newcommand{\smallfrac}[2]{\mbox{$\frac{#1}{#2}$}}
\newcommand{\ket}[1]{| {#1} \ra}
\newcommand{\bra}[1]{\la {#1} |}
\newcommand{\pfpx}[2]{\frac{\partial #1}{\partial #2}}
\newcommand{\dfdx}[2]{\frac{d #1}{d #2}}
\newcommand{\half}{\smallfrac{1}{2}}
\newcommand{\s}{{\mathcal S}}
\newcommand{\jord}{\color{red}}
\newcommand{\kurt}{\color{blue}}
\newtheorem{theo}{Theorem} \newtheorem{lemma}{Lemma}

\title{Ability of Markovian Master Equations to Model Quantum Computers \\ and Other Systems Under Broadband Control}

\author{Gavin McCauley}
\affiliation{U.S. Army Research Laboratory, Computational and Information Sciences Directorate, Adelphi, Maryland 20783, USA} 
\author{Benjamin Cruikshank} 
\affiliation{U.S. Army Research Laboratory, Computational and Information Sciences Directorate, Adelphi, Maryland 20783, USA} 
\affiliation{Department of Physics, University of Massachusetts at Boston, Boston, MA 02125, USA} 
\author{Siddhartha Santra} 
\affiliation{U.S. Army Research Laboratory, Computational and Information Sciences Directorate, Adelphi, Maryland 20783, USA}
\author{Kurt Jacobs}
\affiliation{U.S. Army Research Laboratory, Computational and Information Sciences Directorate, Adelphi, Maryland 20783, USA} 
\affiliation{Department of Physics, University of Massachusetts at Boston, Boston, MA 02125, USA} 
\affiliation{Hearne Institute for Theoretical Physics, Louisiana State University, Baton Rouge, LA 70803, USA} 

\begin{abstract} 
Most future quantum devices, including quantum computers, require control that is broadband, meaning that the rate of change of the time-dependent Hamiltonian  is as fast or faster than the dynamics it generates. In many areas of quantum physics, including quantum technology, one must include dissipation and decoherence induced by the  environment. While Markovian master equations provide the only really efficient way to model these effects, these master equations are derived for constant Hamiltonians (or those with a discrete set of well-defined frequencies). In 2006, Alicky, Lidar, and Zanardi [Phys.\ Rev.\ A \textbf{73}, 052311 (2006)] provided detailed qualitative arguments that Markovian master equations could not describe systems under broadband control. Despite apparently broad acceptance of these arguments, such master equations are routinely used to model precisely these systems. This odd state of affairs is likely due to a lack of quantitative results. Here we perform exact simulations of two- and three-level systems coupled to an oscillator bath to obtain quantitative results. 
Although we confirm that in general Markovian master equations cannot predict the effects of damping under broadband control, we find that there is a widely applicable regime in which they can. Master equations are accurate for weak damping if both the Rabi frequencies and bandwidth of the control are significantly smaller than the system's transition frequencies. They also remain accurate if the bandwidth of control is as large as the frequency of the driven transition so long as this bandwidth does not overlap other transitions. Master equations are thus able to provide accurate descriptions of many quantum information processing protocols for atomic systems.

\end{abstract} 


\maketitle 

\section{Introduction} 

Quantum systems that are subjected to noise and relaxation processes due to an interaction with their environments are referred to as being \textit{open}. Modelling these systems, especially those that are weakly damped, is important in the development of future quantum technologies~\cite{mikeandike, Giovannetti11, Komar14, Jacobs14, WM10}. A system is weakly damped if the damping rates induced by the environment are very small compared to its transition frequencies. Weakly-damped systems can be accurately modeled, in many cases, using a very simple \textit{Markovian master equation} (MME)~\cite{McCauley19x1, Jacobs14, Breuer07, Gardiner10, WM10, Hall14}. The MME is extremely efficient because the model requires no additional degrees of freedom beyond those of the system itself. (While remarkable techniques have now been developed to exactly simulate open quantum systems beyond the regime of master equations, these techniques incur a high numerical overhead~\cite{Bulla03, Chin10, Prior10}.)

When quantum systems are subjected to broadband control, meaning that their Hamiltonians change on a timescale that is similar to their dynamics, in general the methods used to derive master equations break down. As argued in detail by Alicki, Lidar, and Zanardi~\cite{Alicki06}, master equations cannot therefore be trusted to model rapidly controlled systems such as quantum computers~\cite{Alicki06,Rivas10, Albash12, Liu14, Avron15}. Nevertheless, MME's are still regularly used to do so, and little in the way of quantitative results on this question have been obtained. It is not known just \textit{how} inaccurate MME's are for modelling time-dependent systems, nor how this accuracy varies with the bandwidth or other characteristics of the control. Further, since the primary use of MME's in quantum technologies is to describe very small errors due to the effects of dissipation, it is worth knowing whether MME's can predict accurately the  \textit{order-of-magnitude} of errors even if they cannot provide the exact values. Here we not only confirm quantitatively that MME's can be highly inaccurate in describing open systems under broadband control, we also show that there is a non-trivial regime of broadband control in which MME's remain accurate, thus opening up a significant class of systems to efficient simulation. This class includes many implementations of single qubit gates~\cite{Haffner08, Neumann10b, Yao12, Koch07, Blais04}.  

There are two parameters that characterise the timescales of control. The first is the magnitude of the control Hamiltonian, $H_{\ms{c}}$, which determines the speed of the evolution induced by the control. The second is the rate at which $H_{\ms{c}}$ is changed with time. We will refer to the former as the \textit{rate} of the control and the latter as its \textit{bandwidth}. Both can potentially affect the accuracy of the master equation. 

\begin{figure}[t] 
\centering
\leavevmode\includegraphics[width=0.85\hsize]{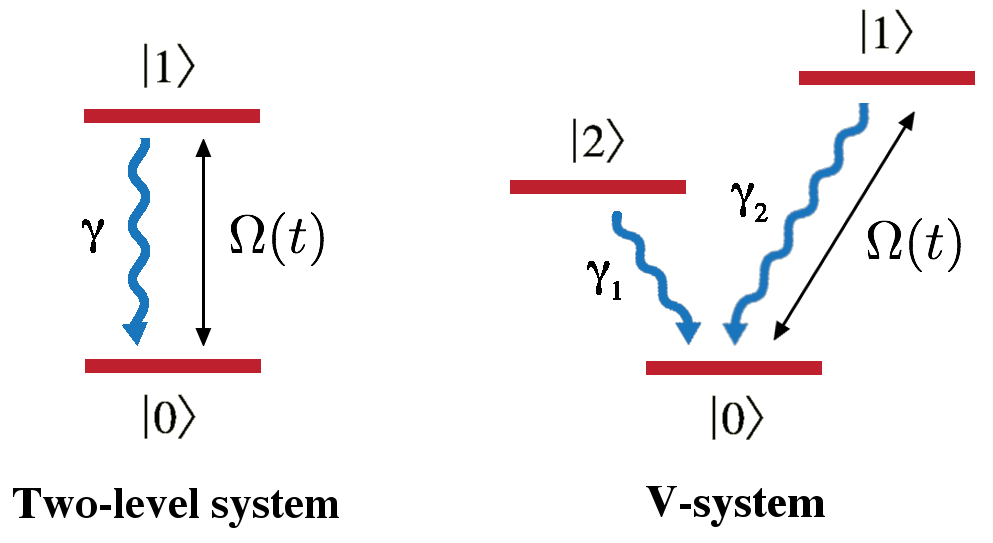}
\caption{(Color online) Diagram of the two- and three-level open systems that we consider. The blue arrows indicate the transitions that undergo damping to a zero-temperature bath. The two-headed arrows indicate the transitions that are driven via a time-dependent Rabi-frequency.}
\label{fig1}
\end{figure}

If both the speed and bandwidth of the control are small compared to the separation of the transition frequencies $\omega_n$, then we can be confident that any driving that the control applies to one transition will not affect the behavior of other transitions. Thus explorations of time-dependent control of a single transition, under this condition, will inform us about multi-transition systems in which separate transitions are controlled independently. By \textit{independent control} we mean that no two driven transitions share a state. Here we explore time-dependent driving of a single transition (by simulating a single qubit) and of a single transition in which a second transition is present but undriven (see Fig.\ref{fig1}). Our results inform about multi-transition systems in the above sense. While we primarily focus here on time-dependent driving, meaning that the control involves coupling the two levels of the transition together, we will also give some results on controlling the frequency of the transition. 

\section{Model of the Environment}

To simulate the exact evolution of an open system we use the standard model of a thermal bath, a continuum of harmonic oscillators. Even though this model is standard, we review it briefly now. Let us denote the energy levels of our open system by $|j\rangle$, so that the Hamiltonian of the system is $H_0 = \sum_{j}E_j|j\rangle$, and denote the upper and lower levels of the $n^{\msi{th}}$ transition by $|k_n\rangle$ and $|l_n\rangle$, respectively. The joint Hamiltonian of the open system and the bath is 
\begin{align} 
    H_{\ms{J}} & = H_0 + H_{\ms{I}} + \int_0^{\Omega_{\ms{c}}} \!\! \hbar \omega b(\omega)^\dagger b(\omega) \, d\omega 
    \label{sysbath}
\end{align} 
where the interaction Hamiltonian is 
\begin{align} 
    H_{\ms{I}} & = \hbar \sum_n g_n (\sigma_n + \sigma_n^\dagger)\int_0^{\Omega_{\ms{c}}} \!\!\!\! \sqrt{J(\omega)} \left[ b(\omega) + b^\dagger(\omega)\right] d\omega  \nonumber
\end{align} 
Here the operator 
\begin{align}
    \sigma_n = |l_n\rangle \langle k_n| 
    \label{sigmas}
\end{align}
is the lowering operator for the $n^{\msi{th}}$ transition, $b(\omega)$ is the annihilation operator for the bath oscillator with frequency $\omega$, and $J(\omega)$ is the spectral density of the oscillator bath. The upper limit $\Omega_{\ms{c}}$ is referred to as the cut-off frequency; the effect of the bath on the open system will only be approximated well by an MME so long as $\Omega_{\ms{c}}$ and the transition frequencies, $\omega_n = (E_{k_n} - E_{l_n})/\hbar$, are much larger than both the damping rates and the Rabi frequencies. When the open system is described well by an MME, the damping rates are given by  
\begin{align}
    \gamma_n & = 2 \pi |g_n|^2 J(\omega_n)  . 
    \label{damping_rates}
\end{align} 
In addition, the energies of the upper levels of the transitions are modified by the Lamb shift. The shift to the energy of level $|k_n\rangle$ is given   by~\cite{Barnett03,McCauley19x1} 
\begin{align} 
    \Delta_n & = |g_n|^2 \mathbb{P} \left[ \int_{-\omega_n}^{\Omega_{\ms{c}}-\omega_n}   \frac{J(\Omega_{\ms{c}} + \omega_n) }{\omega}  d\omega \right] , \nn 
\end{align}
in which $\mathbb{P} \left[\cdot \right]$ denotes the \textit{principle value} of the integral. Since the spectral density affects only the damping rates and Lamb shifts the choice of this density is not critical. We use here the flat density 
\begin{align} 
    J(\omega) = \frac{1}{\Omega_{\ms{c}}} , 
\end{align}
for which   
\begin{align}
    \gamma_n & = 2 \pi \frac{|g_n|^2}{\Omega}   \\
    \Delta_n & = \frac{\gamma_n}{2\pi} \ln\left[\frac{\Omega_{\ms{c}}}{\omega_n}-1 \right] . 
    \label{gamDel}
\end{align} 

Exact simulations of the system and the bath are enabled by a remarkable method developed by Bulla~\textit{et al.\ }\cite{Bulla03} and refined by Chin~\textit{et al.\ }\cite{Chin10, Prior10}, using the matrix-product-state (MPS) method of Vidal~\cite{Vidal03}. We give further details of this method in Appendix~\ref{nummeth}

\section{The Master Equation and the ``adiabatic" extension}

Here we will consider weakly-damped quantum systems with non-degenerate transitions. This means that in addition to the weak damping requirement described above, the differences between the frequencies of any two transitions are also much larger than the damping rates. Systems coupled to a bath as in Eq.(\ref{sysbath}), and that are weakly-damped with $N$ non-degenerate transitions, obey the MME~\cite{McCauley19x1, Jacobs14, Breuer07}
\begin{align}    
  \dot{\rho} = & -\frac{i}{\hbar} \biggl[ H_0 + H_{\ms{L}},\rho \biggr] + \sum_{n=0}^N \gamma_n  \mathcal{D}[\sigma_n] \rho .  
  \label{mme}
\end{align} 
Here we have defined 
\begin{align}
    \mathcal{D}[c] \rho \equiv c \rho c^\dagger - \left(c^\dagger c \rho + \rho c^\dagger c \right)/2 
\end{align}
for an arbitrary operator $c$, the Lamb shift Hamiltonian is 
\begin{align}
    H_{\ms{L}} = \hbar \sum_{n} \Delta_n |k_n\rangle\langle k_n| , 
\end{align}
the ``transition operators" $\sigma_n$ are those defined in Eq.(\ref{sigmas}), and $\gamma_n$ and $\Delta_n$ are as given in Eq.(\ref{gamDel}). 

Our purpose here is to examine under what conditions the evolution of the MME above deviates from that given by the Hamiltonian in Eq.(\ref{sysbath}) when the system Hamiltonian $H$ is time-dependent. More precisely, we will split the system Hamiltonian, now denoted by $H(t)$, into two parts, 
\begin{align}
    H(t) = H_0 + H_{\ms{c}}(t) . 
\end{align}
in which $H_{\ms{c}}(t)$ is the applied time-dependent control. The full model of the system and bath is now given by Eq.(\ref{sysbath}) but with the constant Hamiltonian $H_0$ replaced by $H(t)$. We ask how well the evolution of this time-dependent open system is described by the MME 
\begin{align}    
  \dot{\rho} = & -\frac{i}{\hbar} \biggl[ H(t) + H_{\ms{L}},\rho \biggr] +  \sum_{n=0}^N \gamma_n \mathcal{D}[\sigma_n] \rho . 
   \label{MEW}
\end{align} 
Note that here the transition operators, $\sigma_n$, are still those defined in Eq.(\ref{sigmas}), and thus defined by the eigenstates of $H_0$. 

We will also examine how well a simple time-dependent extension of the MME reproduces the evolution of Eq.(\ref{sysbath}). This time-dependent extension is usually referred to as the \textit{adiabatic master equation} (AME)~\cite{Childs01, Sarandy05}. To obtain the AME we note that the operators that appear in the MME as derived from the system/bath interaction are defined by the eigenstates of the (time independent) system Hamiltonian $H_0$. Thus if the system Hamiltonian changes with time sufficiently slowly, one can expect that at each time the evolution of the system will be given approximately by the MME corresponding to the system Hamiltonian at that time. We thus construct the AME by replacing in the MME the eigenstates of $H_0$ by the (now time-dependent) eigenstates of $H(t)$. Thus the AME is given by 
\begin{align}    
  \dot{\rho} = & -\frac{i}{\hbar} \biggl[ H(t) + \tilde{H}_{\ms{L}}(t),\rho \biggr] + \sum_{n=0}^N \tilde{\gamma}_n(t) \mathcal{D}[\tilde{\sigma}_n(t)] \rho 
\end{align} 
in which 
\begin{align}    
   H(t) & = \sum_j \tilde{E}_j(t) |\tilde{j}(t)\rangle \langle\tilde{j}(t)| , \\  
     \tilde{H}_{\ms{L}}(t) & =  \sum_{n} \hbar \tilde{\Delta}_n(t)  |\tilde{k}_n(t)\rangle\langle \tilde{k}_n(t)| , \\ 
  \tilde{\sigma}_n(t) & = |\tilde{l}_n(t)\rangle \langle\tilde{k}_n(t)| 
\end{align} 
and   
\begin{align} 
    \tilde{\omega}_n & = (\tilde{E}_{k_n}(t) - \tilde{E}_{l_n}(t))/\hbar , \\
    \tilde{\gamma}_n(t) & = 2 \pi \frac{|\tilde{g}_n|^2}{\Omega} , \\
    \tilde{\Delta}_n(t) & = \frac{\tilde{\gamma}_n}{2\pi} \ln\left[\frac{\Omega_{\ms{c}}}{\tilde{\omega}_n}-1 \right] , 
\end{align} 
where  
\begin{align}
    \tilde{g}_n & = \langle\tilde{l}_n(t)|  \biggl[ \sum_m (\sigma_m + \sigma_m^\dagger) \biggr] |\tilde{k}_n(t)\rangle . 
\end{align}

\textit{Note regarding the Lamb shift:} For our simulations we set $\Omega_{\ms{c}} = 10\omega$, with the result that the Lamb shift is $\Delta \approx \gamma/3$. It turns out that at this value the Lamb shift has a negligible effect on the dynamics, at least for our purposes. The numerical values we use for $\omega$ are between $2\pi$ and $8\pi$, with all except those in Fig.4 using $\omega = 8\pi$. Taking $\omega = 8\pi$ and the largest value we use for $\gamma$, which is $\gamma = 5\times 10^{-3}$, the resulting Lamb shift is $\Delta \approx 2\times 10^{-4}$. We find that the error in the evolution induced by this Lamb shift is about $5\times 10^{-6}$. For this value of $\gamma$ the effect on the relative error calculation is therefore on the order of $10^{-3}$, which is below our numerical error threshold (see below). 

\section{Measure of Accuracy}

To explore the accuracy of master equations we need to define an appropriate measure of this accuracy. The purpose of the master equations is to evaluate the effect of the bath on the system. For quantum technologies the appropriate regime is that in which the damping rates are very slow compared to the timescale of interest, and thus the effect of the damping and associated decoherence is small. We want to know how well the master equation faithfully reproduces this small effect. Thus in quantifying how well the master equation performs, it is sensible to report this accuracy as a fraction of the total effect of the damping. 

If the bath modifies the evolution of a system by $\Delta_0$, and the MME predicts that the evolution is modified instead by $\Delta_1$, then we will report the error incurred by the master equation as 
\begin{align}
    \epsilon = |\Delta_1 - \Delta_0|/|\Delta_0|, 
\end{align}
and refer to $\epsilon$ as the \textit{relative error}. Let us denote the evolution of the upper level of the driven transition for i) the undamped system, ii) the system coupled to the bath, and iii) that predicted by an MME by $p_{\ms{e}}^{H}$, $p_{\ms{e}}^{\ms{exact}}$, and $p_{\ms{e}}^{\ms{ME}}$, respectively. We will quantify the effect of the bath on the system by 
\begin{align}
    \Delta_0(T) = \int_0^T \! |p_{\ms{e}}^{\ms{H}}(t) - p_{\ms{e}}^{\ms{exact}}(t)| \, dt , 
\end{align}
and similarly the effect predicted by the MME as 
\begin{align}
    \Delta_1(T) = \int_0^T \! |p_{\ms{e}}^{\ms{H}}(t) - p_{\ms{e}}^{\ms{ME}}(t)| \, dt . 
\end{align}


\section{Classes of Control}

Consider now the general problem of determining the ability of an MME to predict the behavior of a time-dependent open quantum system with $N$ non-degenerate transitions. There are a number of fairly distinct ways in which the system may be made time-dependent: one might change the energy levels, the eigenstates, or apply a coupling between the eigenstates that, while modifying them only a little, causes population to oscillate between them. Note that these forms of time-dependence are not fully distinct because as the coupling (the Rabi frequency) between two eigenstates increases, and the frequency of this coupling decreases, what may be considered as a driving term turns into a control that slowly varies the energy levels and eigenstates. Nevertheless, we anticipate that breaking control into these ``types'' will be of some use because they may have different effects on the accuracy of the master equations. 

Driving of a transition with lower level $|j\rangle$ and upper level $|k\rangle$ takes the form 
\begin{align}
    H_{\ms{d}}(t) = \hbar \Omega(t)\cos(\omega_{\ms{d}} t)\left(  \cos[\theta(t)]\sigma_x + \sin[\theta(t)]\sigma_y \right) 
    \label{Hdrive}
\end{align}
$\sigma_x = \sigma + \sigma^\dagger$ and $\sigma_y = i( \sigma - \sigma^\dagger)$, in which $\sigma = |j\rangle \langle k|$, are the Pauli spin operators. The amplitude of the coupling, which we denote here by $\Omega(t)$, is called the Rabi frequency as it gives the frequency at which the coupling flips population between the two levels. The angle $\theta$ is the phase of the coupling, which we will call the Rabi angle. We have multiplied the entire expression by $\cos(\omega_{\ms{d}} t)$ so that the spectrum of the driving is centered at $\omega_d$. The Hamiltonian of the two-level system, including the driving, is 
\begin{align}
   H_{\ms{c}}(t) =  \hbar \left(\frac{\omega}{2}\right) \sigma_z +  H_{\ms{d}}(t) . 
\end{align}
in which $\sigma_z = |k\rangle\langle k| -  |j\rangle\langle j|$. If $\omega_d = \omega$, and $\Omega$ and $\theta$ are constant, then the driving is resonant with the transition. 

We will not consider the factor $\cos(\omega t)$ as being part of the time-dependence of $H_{\ms{d}}$; under the rotating-wave approximation, and in the interaction picture, the time-dependence of this term vanishes. Further, because the factor $\cos(\omega t)$ has a fixed frequency, a master equation that is more accurate than Eq.(\ref{mme}) could be derived for it by taking this frequency into account. As such, we are not concerned with this kind of time-dependence here since it is not beyond the reach of present methods. Because of this we refer to $H_{\ms{d}}$ as time-dependent control only if $\Omega$ or $\theta$ are time-dependent, and we define the bandwidth of the control as the maximum frequency component of $\Omega$ and $\theta$. We will also consider control in which the Rab frequency is fixed while the frequency of the transition is changed with time.  

\begin{figure}[t] 
\centering
\leavevmode\includegraphics[width=1\hsize]{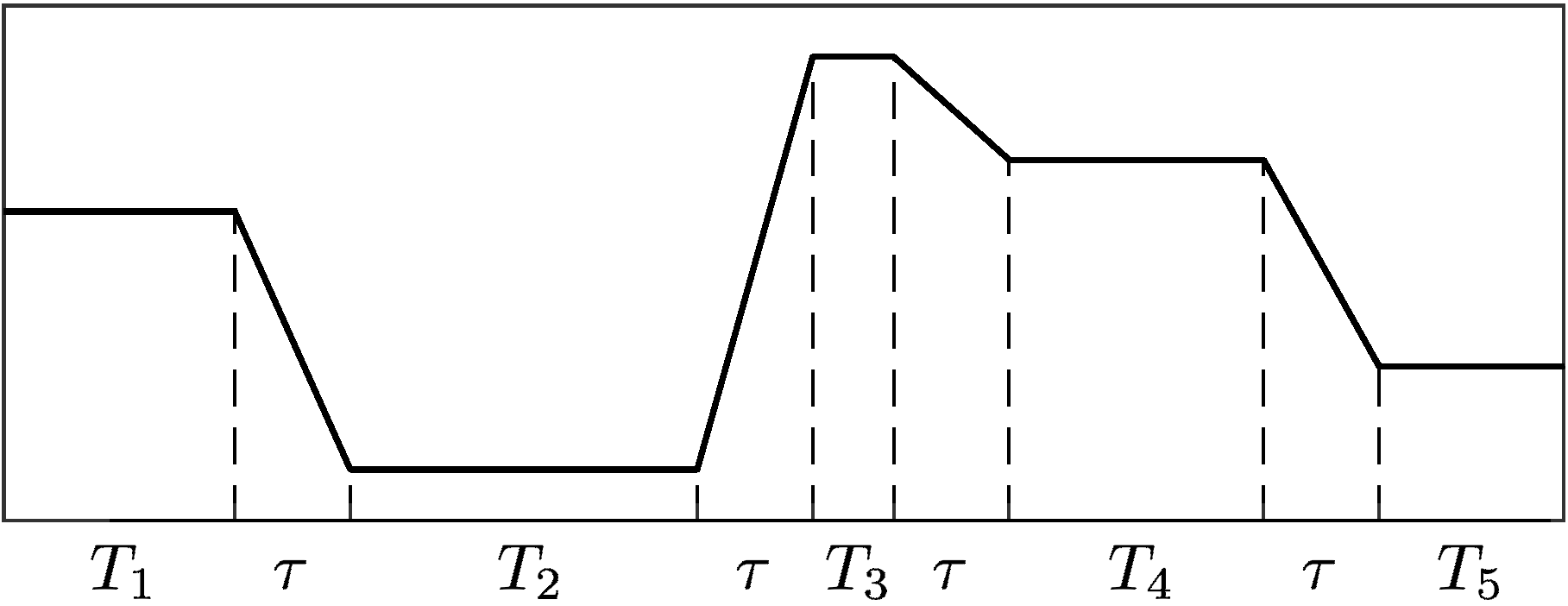}
\caption{The class of the piecewise control functions that we use. On all the odd intervals the functions are constant, and on even intervals they are linear. The linear intervals connect the values on the adjoining intervals. The duration of the odd intervals varies, while the even intervals all have the same duration, $\tau$. By reducing $\tau$ we increase the maximum rate of change of the control function. When $\tau=0$ the functions are piecewise-constant. 
}
\label{fig2}
\end{figure}

\subsection{Two forms for time-dependence} 

We will explore control in which the time-dependence of the parameters, namely $\Omega$, $\theta$, and $\omega$, is characterized in two distinct ways. In the first we represent the time-dependent functions by a  Fourier series with a finite number of terms:
\begin{align}
    f(t) = c_0 + \sum_{k=1}^K c_k \cos(k \nu t) + s_k \sin(k \nu t) 
    \label{Fseries}
\end{align}
This form allows us to consider controls with a specific and adjustable bandwidth (this bandwith is $\omega_{\ms{w}} = K\nu$). We can randomly sample the space of these functions by choosing the coefficients of the Fourier series to be independent random variables. 

The second form we consider for the time-dependent control functions has a well-defined maximum rate of change. We define these in a piecewise fashion as depicted in Fig.\ref{fig2}. Dividing time into segments, and labelling the segments as $1, 2, \ldots, N$, the functions are constant on odd segments. Between these segments the function is linear so as to connect the values on the odd segments. The duration of the linear segments are all identical, and equal to $\tau$. As $\tau \rightarrow\infty$ the functions become discontinuous and are piecewise constant. We can adjust the maximum rate of change of a given function simply by changing $\tau$.   

\begin{figure}[t] 
\centering
\leavevmode\includegraphics[width=1\hsize]{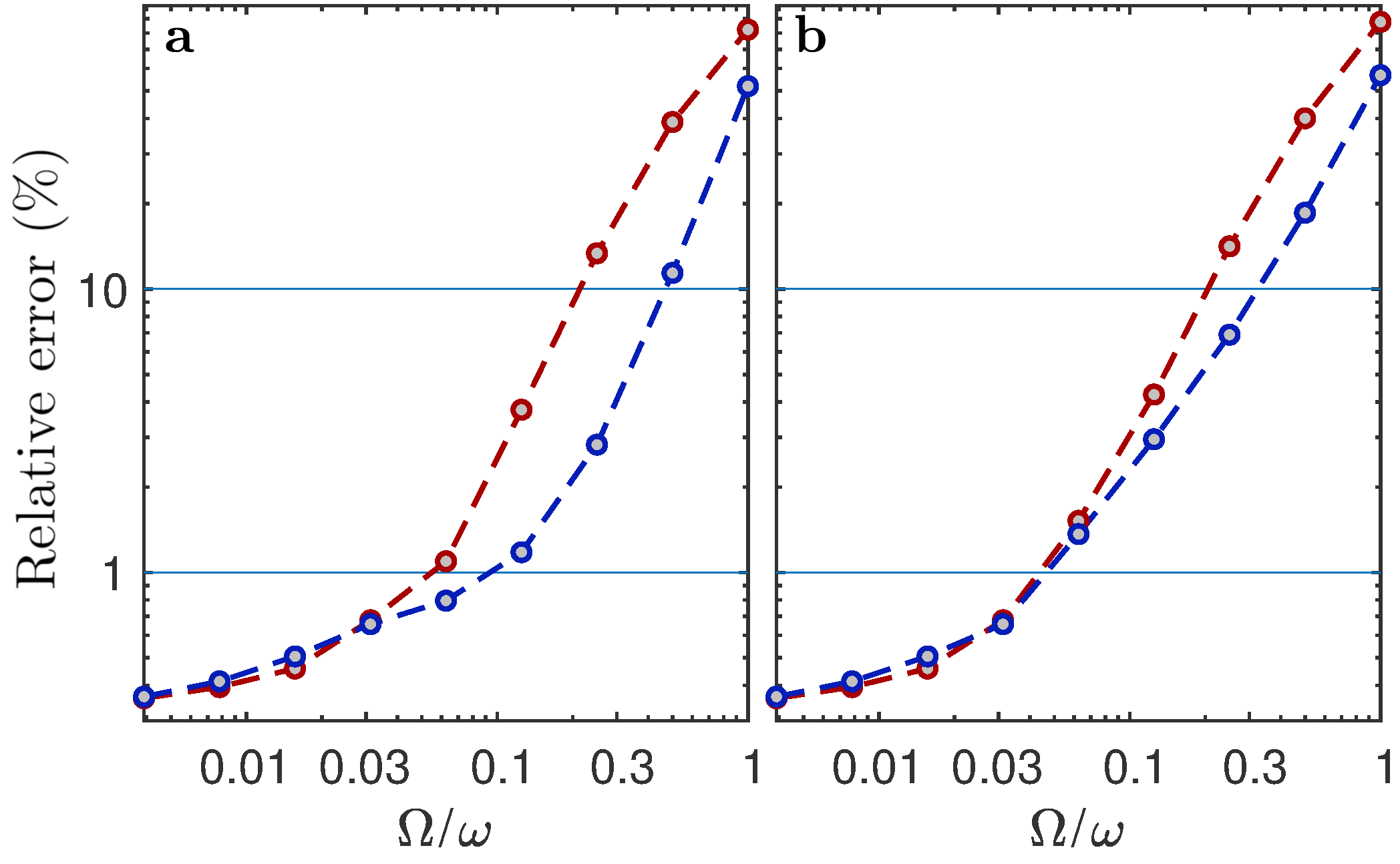}
\caption{(Color online) The error of the Markovian master equation (MME), Eq.(\ref{MEW}), and its adiabatic time-dependent version (AME), in simulating a weakly-damped, driven two-level system with Rabi frequency $\Omega$. The damping rate is $\gamma = 10^{-3}\omega/(8\pi)$ where $\omega$ is the transition frequency. The bath has a uniform spectral density and a cut-off frequency $\Omega_{\ms{c}} = 10\omega$. Blue circles: MME; red circles: AME. The dashed lines merely connect the circles. (a) Relative error, $\epsilon$, for the fixed duration $T = 2\pi/\Omega_{\ms{min}} = 8\times 10^{-3}/\gamma$; (b) Relative error for the duration $T(\Omega) = 2\pi/\Omega$. 
}
\label{fig3}
\end{figure}

\section{Results}

Before we explore the accuracy of master equations for time-dependent control, it is useful to examine their accuracy for constant driving to serve as a reference point. In Fig.\ref{fig3} we plot the accuracy of the MME and AME for a two-level system driven on resonance with a constant Rabi frequency, $\Omega$, as a function of this frequency. For all our simulations the cut-off frequency is set at $\Omega_{\ms{c}} = 10\,\omega$, where as usual $\omega$ is the frequency of the transition. For Fig.\ref{fig3} the damping rate is $\gamma = \omega/(8\pi)\times 10^{-3}$ and the smallest value of the Rabi frequency, $\Omega$, is $\Omega_{\ms{min}} = \omega/32$. 

In Fig.\ref{fig3}a we show the relative error for the MME (blue) and AME (red) for a fixed evolution time of $T = 2\pi/\Omega_{\ms{min}} = 8\times 10^{-3}/\gamma$. For the purposes of quantum information processing, it is usually the error per logical operation (or ``gate") that is relevant, and the gate time is usually on the order of the inverse Rabi frequency. Because of this, in Fig.\ref{fig3}b we show the relative error, again as a function of $\Omega$, this time for an evolution time of $T = 2\pi/\Omega$. 

We see from Fig.\ref{fig3} that so long as $\Omega \leq \omega/10$ the error in the ME is no more than $1\%$. We also note that as we reduce the Rabi frequency there is point a which the relative error stops decreasing. This is to be expected: there is an ``error floor'' for the MME due to the finite ratios of $\gamma/\omega$ and $\omega/\Omega_{\ms{c}}$; the MME is only asymptotically exact as these ratios tend to zero~\cite{Santra17}. In Appendix~\ref{numacc} we discuss further details regarding the origin of the error floor, including numerical accuracy. 

In Fig.\ref{fig7}c we explore what happens for a constant drive at a fairly large Rabi frequency ($\Omega = \omega/4$) when the driving is off-resonant. Interestingly the behavior of the MME and AME are quite different in this case. For the MME the error drop dramatically for driving frequencies much lower than the transition frequency, and for the AME it drops dramatically when the driving frequency is much higher than the transition frequency. 

\begin{figure}[b] 
\centering
\leavevmode\includegraphics[width=1\hsize]{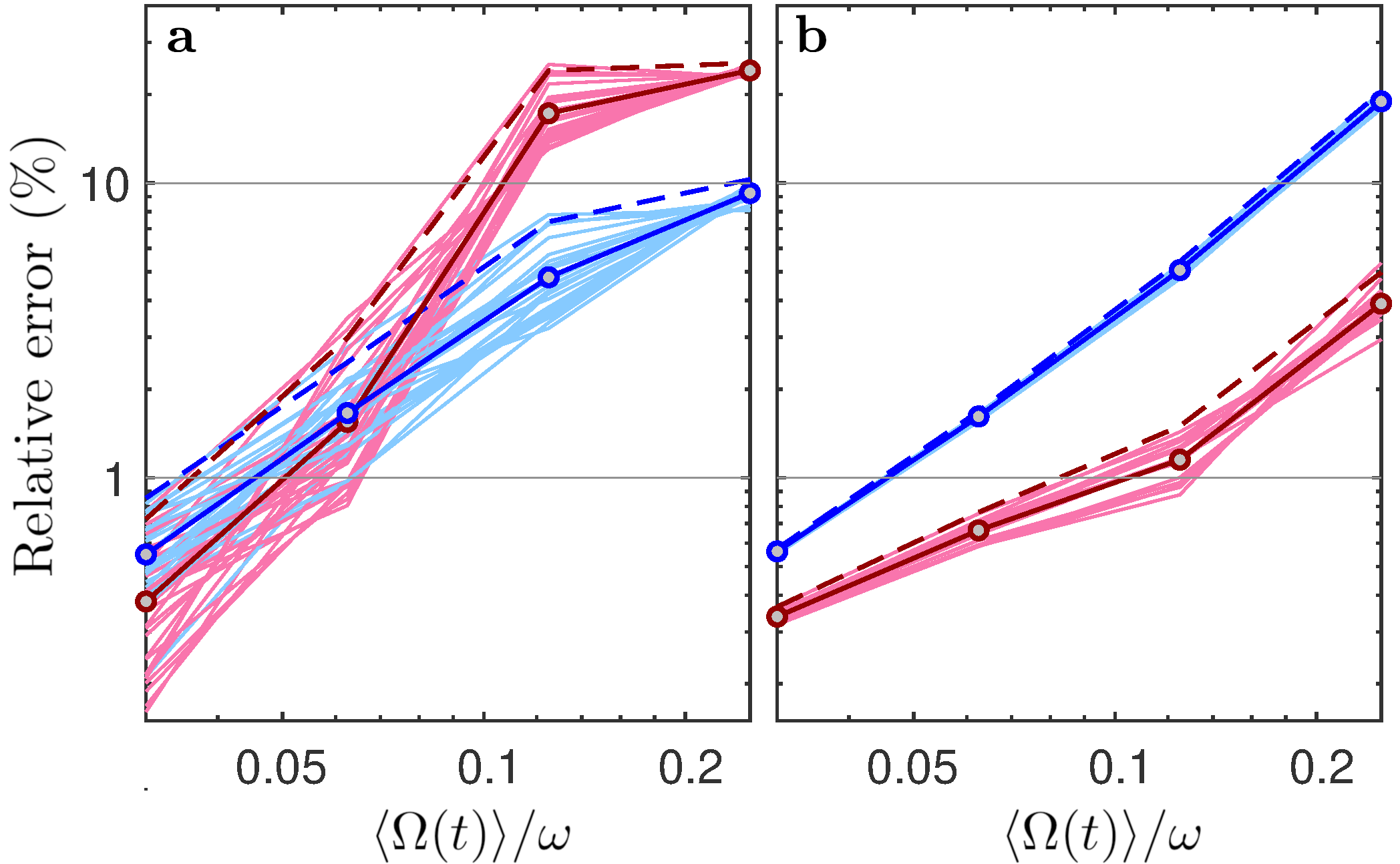}
\caption{(Color online) The error of the Markovian master equation (MME), Eq.(\ref{MEW}), and its adiabatic version (AME), in simulating a weakly-damped, driven two-level system with transition frequency $\omega$, when the Rabi frequency is modulated arbitrarily within a given bandwidth $\omega_{\ms{w}}$. a) $\omega_{\ms{w}} = \omega/8$; b) $\omega_{\ms{w}} = \omega$. The parameters are those used in Fig.~\ref{fig3}(a). Light blue: error of the MME for 28 samples of the time-dependent drive; Light red: error of the AME for 20 samples. Dark lines: average error; Dashed lines: average error plus two standard deviations.}
\label{fig4}
\end{figure}

We now turn to time-dependent driving of a two-level system, in which the drive has a non-zero bandwidth centered at the transition frequency. Here we set $\theta=0$ and choose for $\Omega(t)$ bandwidths of $\omega/8$ and $\omega$. To represent $\Omega(t)$ we use the Fourier series of Eq.(\ref{Fseries}) with $K=20$. Thus $\nu$ is equal to $\omega/160$ and $\omega/20$ for bandwidths of $\omega/8$ and $\omega$, respectively. Since different signals with the same bandwidth may lead to different relative errors, we sample different Fourier series by sampling the coefficients $c_k$ and $s_k$ as independent Gaussian random variables with unit variance and zero mean. Since we want to see how the relative error of the master equations changes as we increase the strength of the drive, we use as our measure of this strength the root mean square Rabi frequency defined as 
\begin{align}
    \Omega_{\ms{rms}}  \equiv \sqrt{\frac{1}{T} \int_0^T \!\!\! \Omega^2(t)\, dt} \, ,  
\end{align} 
in which $T$ is the duration of the evolution. To obtain sample functions for $\Omega(t)$ that have specified values for $\Omega_{\ms{rms}}$ we sample the coeficients $c_k$ and $s_k$ independently and merely scale the resulting $f(t)$ to obtain the desired value of $\langle \Omega \rangle$. For a given sample of $f(t)$ we calculate the relative error for different values of $\langle \Omega \rangle$ obtained by scaling $f(t)$. 

In Fig.\ref{fig4}(a) we display the relative errors for a set of 28 sample Fourier series with a bandwidth of $\omega_{\ms{w}} = \omega/8$. As in Fig.\ref{fig3}(b) We calculate the relative error for a duration, $T =  2\pi/\Omega_{\ms{rms}}$, that decreases with the rms Rabi frequency. We plot the error for each sample function for a range of values of $\langle \Omega \rangle/\omega$. We also plot the mean of the samples (as a line with circles) and this mean plus two standard deviations (as a dashed line). In Fig.\ref{fig4}(b) we plot the errors for a set of 20 samples for functions with a bandwidth of $\omega_{\ms{w}} = \omega$. The first thing we see is that, as for constant driving, the relative error increases with the strength of the drive. For both bandwidths the error is larger than that for a constant drive, as we might expect. Two very notable differences between the relative errors for the two bandwidths are that for the larger bandwidth the spread in the errors is much less, and the adiabatic master equation does a much better job, than for the smaller bandwidth. 

The main conclusion we can draw from Fig.\ref{fig4} is that for a two-level system, even for driving that has a bandwidth as large as the transition frequency, so long as the rms Rabi frequency is sufficiently small compared to the transition frequency, the MME and AME provide good models of damping. For relatively large Rabi frequencies, however, both master equations break down. 

\begin{figure}[t] 
\centering
\leavevmode\includegraphics[width=1\hsize]{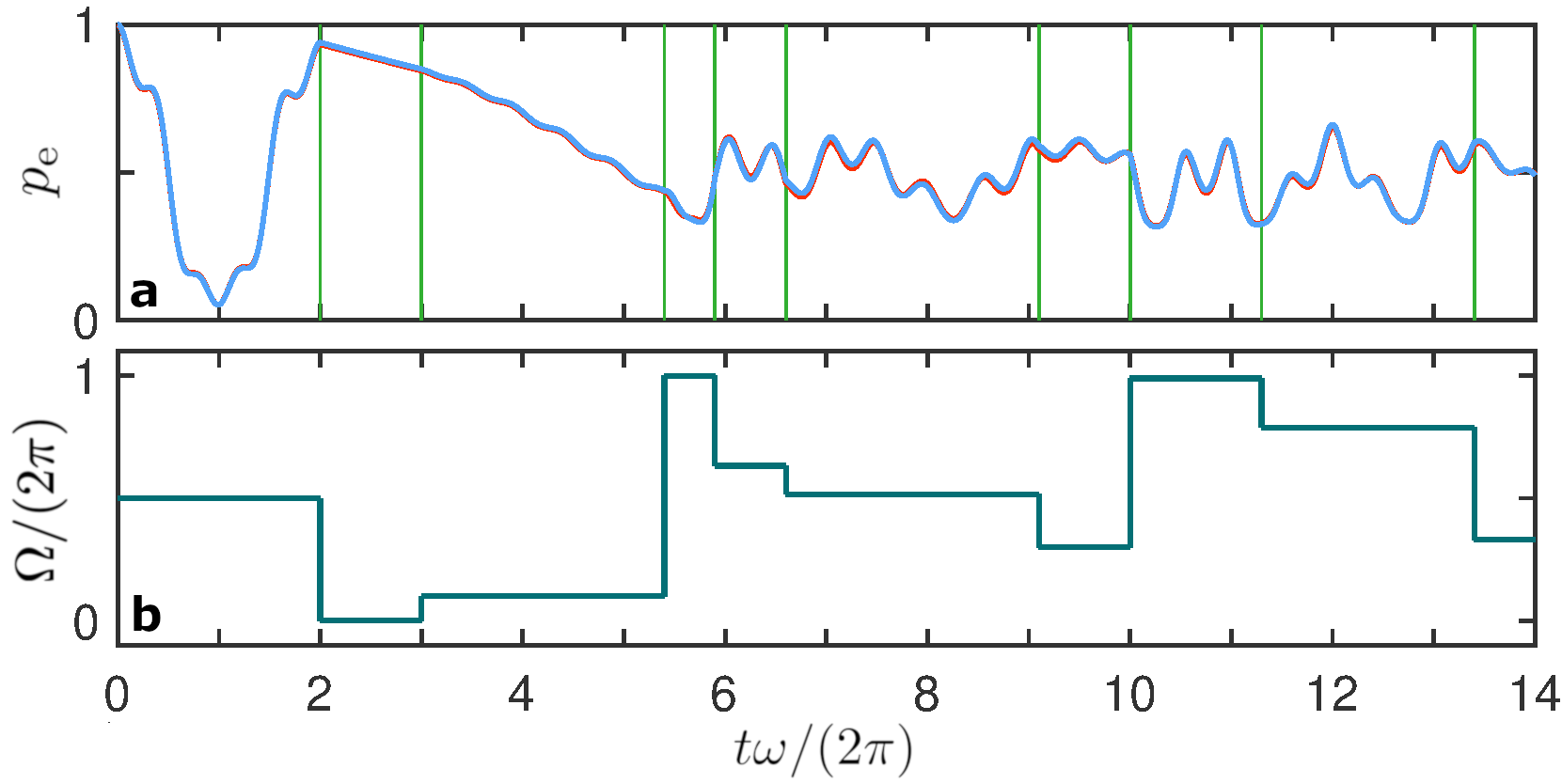}
\caption{(Color online) a) Population of the  excited state of a two-level system under a control protocol for which the magnitude of the time-dependent Rabi frequency is shown in b). The decay rate is $\gamma = \omega/100$ with $\omega$ the transition frequency.(Here we use a larger decay rate  so that its effect is perceptible on the plot.)  blue: exact simulation; red: Markovian master equation (Eq.(\ref{MEW})).}
\label{fig5} 
\end{figure}

We now consider control signals of the piecewise form depicted in Fig.\ref{fig2}. We examine first an example control function in which $\tau=0$, with the result that it represents instant switching between a sequence of constant resonant drives. We choose the example control function so that the sequence of constant drives, defined by values for $\Omega$ and $\theta$ in Eq.(\ref{Hdrive}), is essentially random. In Fig.\ref{fig5}b we show the Rabi frequency as a function of time for this example, and in Fig.\ref{fig5}a we show the resulting evolution of the excited state population with and without damping at the rate $\gamma = \omega/100$. The average Rabi frequency for the control function is $\langle \Omega(t)\rangle \approx \omega_0/2.3$ and the average inverse length of the intervals (the average switching rate) is $\approx (3/4)\omega_0/(2 \pi)$.

\begin{figure}[t] 
\centering
\leavevmode\includegraphics[width=1\hsize]{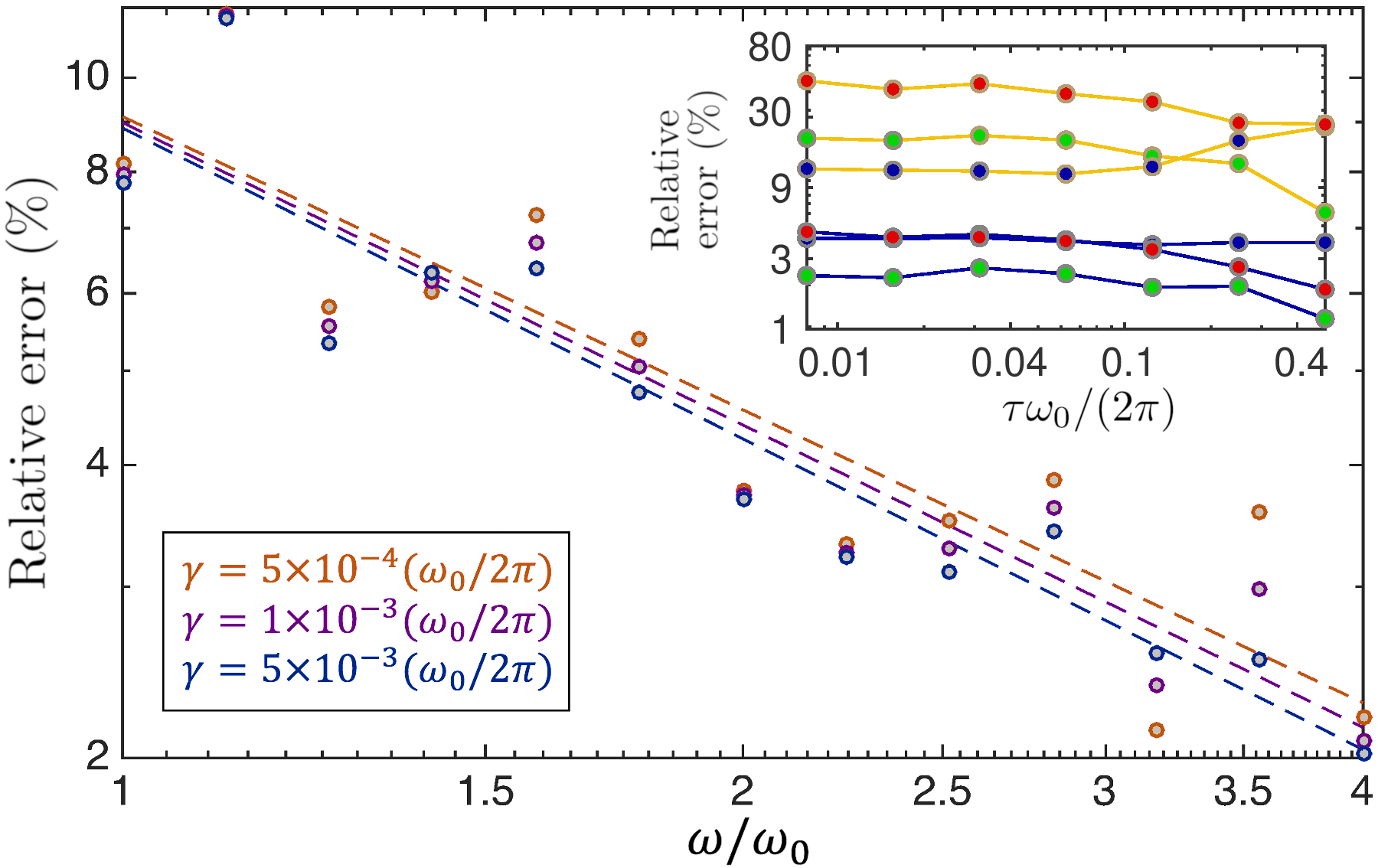}
\caption{(Color online) The error incurred by the MME in simulating a transition in which the Rabi frequency, $\Omega(t)$, is switched instantaneously at a discrete set of times. In this plot the control protocol (the function $\Omega(t)$) remains fixed and the resulting error is shown as a function of the transition frequency, $\omega$, as well as for three values of the damping rate $\gamma$. The dashed lines are the straight lines of least-squares best-fit for each of the values of $\gamma$. For comparison with Figs.\ref{fig3} and \ref{fig4}, in this plot $\langle \Omega(t)\rangle/\omega$ varies from $0.43$ on the left down to $0.11$ on the right. \textit{Inset:} The error for three sequences as a function of the switching time, $\tau$, with (yellow) and without (blue) the presence of a third level that forms a second transition with frequency $\omega_0/2$. The circles (red,green blue) distinguish the three control sequences. }
\label{fig6}
\end{figure}

For the above control function, we examine how the accuracy of the master equations changes as the transition frequency and cut-off frequency are increased. The cut-off frequency is set at $\Omega_{\ms{c}} = 10 \omega$. In Fig.\ref{fig6} we present the results for three values of the damping rate. For the purposes of this plot we have defined a fixed reference frequency $\omega_0$, and vary $\omega$ over the range $[\omega_0,4\omega_0]$. While we are limited in the range of transition frequencies that we can practically simulate, for all three values of the damping there is a clear downward trend as $\omega$ increases. The lines of best-fit indicate that the relative error decreases as $1/\omega$ (the slopes of all three lines are equal to unity to within 0.1\%). This scaling is what we would expect for a constant drive. There are also large fluctuations in the relative error as $\omega$ is changed, which is to be expected since the dynamics under the driving changes dramatically with $\omega$. Note that unlike Figs.\ref{fig1} and \ref{fig2}, here the ratios $\gamma/\omega$ and $\gamma/\Omega_{\ms{c}}$ are not fixed but scale as $1/\omega$, so that there is no floor on the error as $\omega$ is increased. 

\begin{figure*}[t] 
\centering
\leavevmode\includegraphics[width=1\hsize]{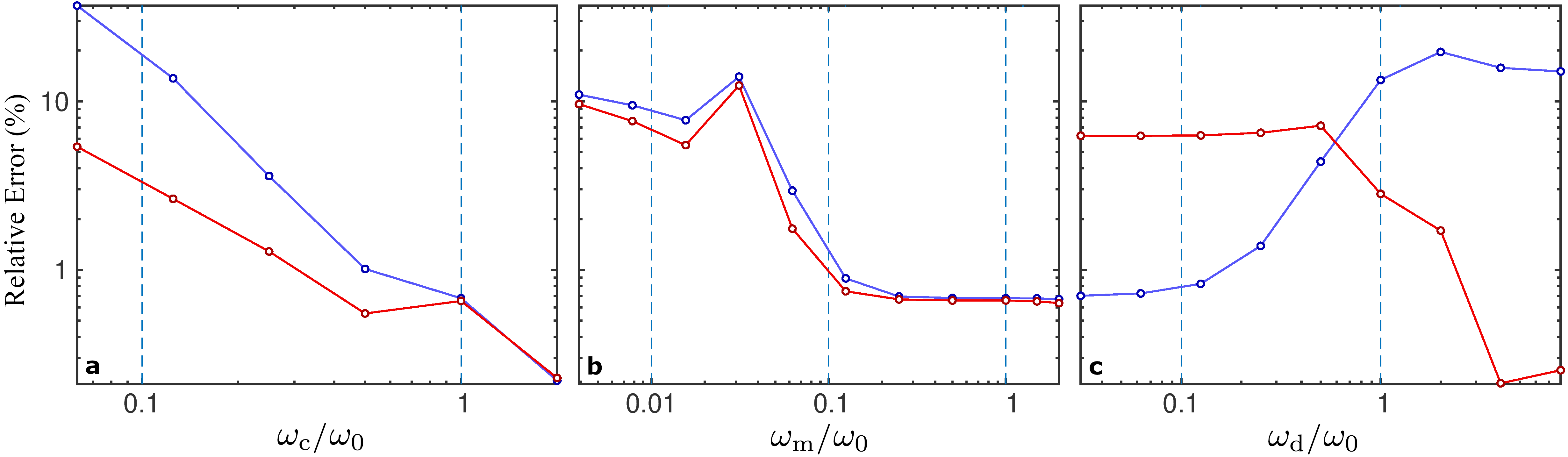}
\caption{(Color online) Relative error of the master equations (MME is blue; AME is red). a) Driving at frequency $\omega_0$ with fixed Rabi frequency $\Omega = \omega_0/32$. The transition frequency is modulated about the value $\omega_{\ms{c}}$, with modulation amplitude $\omega_{\ms{rms}} \approx \omega_0/60$, center frequency $\omega_{\ms{m}} = \omega_0/2$ and bandwidth $\omega_{\ms{w}} = \omega_0/2$. b) The driving is the same as in a) and the transition frequency is modulated about the value $\omega_0$ at a single frequency $\omega_{\ms{m}}$ with modulation amplitude $\omega_0/32$. c) The transition and Rabi frequencies are fixed at $\omega_0$ and $\omega_0/4$, respectively, and the driving frequency is scanned.}
\label{fig7}
\end{figure*}

From Fig.\ref{fig6} we see that the relative error at a transition frequency of $\omega = 4\omega_0 \approx 9 \langle \Omega(t)\rangle$ is about $2\%$, and given the trend we can expect that it will continue to decrease with further increases in $\omega$. Note that the master equations achieve this accuracy even though the bandwidth of the control is infinite. 

We also examine what happens as the switching time $\tau$ is increased so that the bandwidth is reduced. In the inset in Fig.\ref{fig6} we plot the error for three sample protocols, including a scaled version of the one displayed in Fig.\ref{fig5}, as a function of the switching time. In this case we set $\omega = 4\omega_0$ and all three control functions are scaled so that they have the same average Rabi frequency, $\langle \Omega(t)\rangle \approx \omega_0/4$. We see that the while the relative errors are somewhat different for each sample control, there is little change in error as $\tau$ in increased. (There is an indication that the error may start to reduce above $\tau \approx 0.2 \times 2\pi/\omega$.)

\subsection{Presence of a third level}

We now add a third level to our two level system. This level has an energy between that of original ground and excited states and decays to the ground state, so that the result is a three-level ``V" system. We continue to drive only the original transition, so as to examine whether the mere presence of this third level has an effect on the dynamics of the driven transition. Let us denote the frequency of the transition between the ground state and the third level by $\omega_2$. We expect that the effect of the third level will be negligible so long as both the Rabi frequency and bandwidth of the drive on the original transition are smaller than $\omega_2$. The key question is whether the MME remains accurate when the bandwidth of the driving is broad enough to overlap the frequencies of other transitions. 

We first confirm that the third level has little effect if $\omega_2$ is greater than the driving and Rabi frequencies, and the bandwidth at which the Rabi frequency is modulated (the control bandwidth). Setting $\omega_2 = \omega/2$, we calculate the relative error under a drive given by $H(t) = \hbar \Omega \cos(\omega_{\ms{m}}) \cos(\omega t) \sigma_x$ for each of the Rabi frequencies $\Omega = \omega/10$, $\omega/33$, $\omega/100$, and modulation frequencies $\omega_{\ms{m}} = \omega/8$, $\omega/16$, $\omega/32$, $\omega/64$. For all these cases the error is less than 1\%, in line with the results for the two-level system shown in Figs.~\ref{fig3} and \ref{fig4}.

We now consider the three sets of piecewise-linear control functions used to obtain the results shown in Fig.\ref{fig6}. The bandwidth of these control functions is set by the parameter $\tau$, and is approximately $2\pi/\tau$. In the inset of Fig.\ref{fig6} we plot the error of the three sample control functions for $\tau = 2\pi/(2^n\omega_0)$ with $n = 1,2,\ldots,8$. The blue lines give the relative error for the two-level system, and the yellow lines this error in the presence of the third level. Note that the average Rabi frequency of these control functions is well below the frequency of the second transition, $\omega_2$, as $\Omega = \omega_0/4 = \omega_2/2$. The range of bandwidths on the other hand is $2^n\omega_0$ with $n=1,2,\ldots,8$. We see from the plots that the presence of the third level dramatically increases the relative error. 

\subsection{Controlling the transition frequency}

We now present some results on the accuracy of the MME and AME for a two-level system when the transition frequency is changed with time. First, we note that if a transition is undriven, so that the decay into the vacuum is its sole dynamics, then changing the transition frequency has essentially no effect on the dynamics and thus no effect on the accuracy of the master equation (at least for the value of the cut-off frequency we use here). 

We now consider what happens when the transition is driven with a drive frequency of $\omega_{\ms{d}} = \omega_0$ and a fixed Rabi frequency of $\Omega = \omega_0/32$, while the transition frequency is modulated by a function with a given bandwidth and RMS amplitude. To this end we choose 
\begin{align}
    \omega(t) = \omega_{\ms{c}} + \sum_{k=-N}^N c_k \cos(\nu k + \omega_{\ms{m}}) + s_k \sin(\nu k + \omega_{\ms{m}})
\end{align}
in which $N$ is an integer. Here $ \omega_{\ms{c}}$ is the ``center'' (also the time-averaged) value of the transition frequency, $\omega_{\ms{m}}$ is the center frequency of the modulation, and $\omega_{\ms{w}} = N\nu/2$ is the modulation bandwidth. The RMS amplitude of the modulation is  
\begin{align}
    \omega_{\ms{rms}} = \sqrt{\frac{1}{2}\sum_k (c_k^2 + s_k^2)}
\end{align} 
We modulate the transition frequency with a bandwidth of  $\omega_{\ms{w}} = \omega_0/2$ around a center frequency of $\omega_{\ms{m}} = \omega_0/2$ with a RMS amplitude of  $\omega_{\ms{rms}} \approx \omega_0/60$. The damping rate is $\gamma = 10^{-3}\omega/(8\pi)$. In Fig.\ref{fig7}a we show the resulting error of the two master equations as a function of the center transition frequency, $\omega_{\ms{c}}$. We see that when the drive is on resonance with the average transition frequency the error is larger than without the modulation (compare with Fig.\ref{fig3}a) but still below 1\%. Interestingly we find that in this case changing the bandwidth (from $\omega_0/64$ to $\omega_0/2$), and the RMS amplitude of the modulation (from $\omega_0/128$ to $2\omega_0$) has very little effect on the relative error. 

In Fig.\ref{fig7}b we examine the error resulting from modulating the transition frequency at a single modulation frequency when the center transition frequency is resonant with the drive. The drive is the same as for Fig.\ref{fig7}a, and this time we scan the frequency of modulation from $\omega_0/256$ to $2\omega$. From the plot we see that for modulation frequencies from $\omega_0/8$ to $2\omega$ the error changes very little and is below $1\%$. For slower modulation frequencies, at least down to $\omega_0/256$, the error becomes much larger for both master equations. Of course, for very slow modulation modulation frequencies the error must again become small. The message here is that if one is modulating the transition frequency, the master equations are only accurate within certain frequency windows, at least for Rabi frequencies above $\omega_0/32$. 

\section{Summary}
 
To summarize our results, we have found, surprisingly, that the MME is able to model broadband control of a single weakly-damped transition with relative errors of only a few percent so long as the Rabi frequency is small compared to the transition frequency. Further, while practical considerations limit the regimes we can explore (particularly the size of the transition and cut-off frequencies), our results indicate that the accuracy of the MME continues to increase as the Rabi frequency is reduced with respect to $\omega$ and $\Omega_{\ms{c}}$. We have also found that this accuracy is only achieved when there are no other transitions within the bandwidth of the controls. One cannot merely drive a system with arbitrary controls and expect the MME to apply without taking into account the control bandwidth and the frequencies of all transitions. 

While we have only considered driving a single transition at a time, we would expect similar results when multiple connected transitions are driven (e.g. in frequency conversion protocols~\cite{Gard17, Vogt19}). Since Rabi frequencies and control bandwidths tend to be much smaller than transition frequencies for implementations of quantum information processing in atomic systems (e.g. cold atoms~\cite{Levine19}, atomic gasses~\cite{Cox18} and ion traps~\cite{Ge19}) Markovian master equations can be expected to provide accurate models for most of these protocols. This is not the case for superconducting platforms where transition frequencies are much lower. However, for superconducting systems it is not clear that master equations are necessarily accurate models even in the absence of time-dependent control because the environment is much more complex. Our results have shown, as one would expect, that even when time-dependent control causes Markovian master equations to lose their accuracy, they still provide correct \textit{order-of-magnitude} estimates for the effects of a zero-temperature bath. For quantum information processing such estimates are often sufficient, since environmental effects are required to be very small. 

Given the results we have presented, we would suggest that authors using master equations to simulate controlled open quantum systems should make it clear whether their results regarding the effects of dissipation can be taken as quantitatively accurate, or merely as order-of-magnitude estimates.  

\section{Acknowledgments} 
KJ is grateful to Jordan Horowitz for helpful discussions and Hannes Pichler for helpful comments on the manuscript. This research was supported in part by appointments to the Student and Postgraduate Research Participation Programs at the U.S Army Research Laboratory~\footnote{These programs are administered by the Oak Ridge Associated Universities (ORAU) through an interagency agreement between the U.S Department of Energy and USARL.}. 


\appendix

\section{Accuracy of the simulations}
\label{numacc}

\begin{figure}[t] 
\centering
\leavevmode\includegraphics[width=1\hsize]{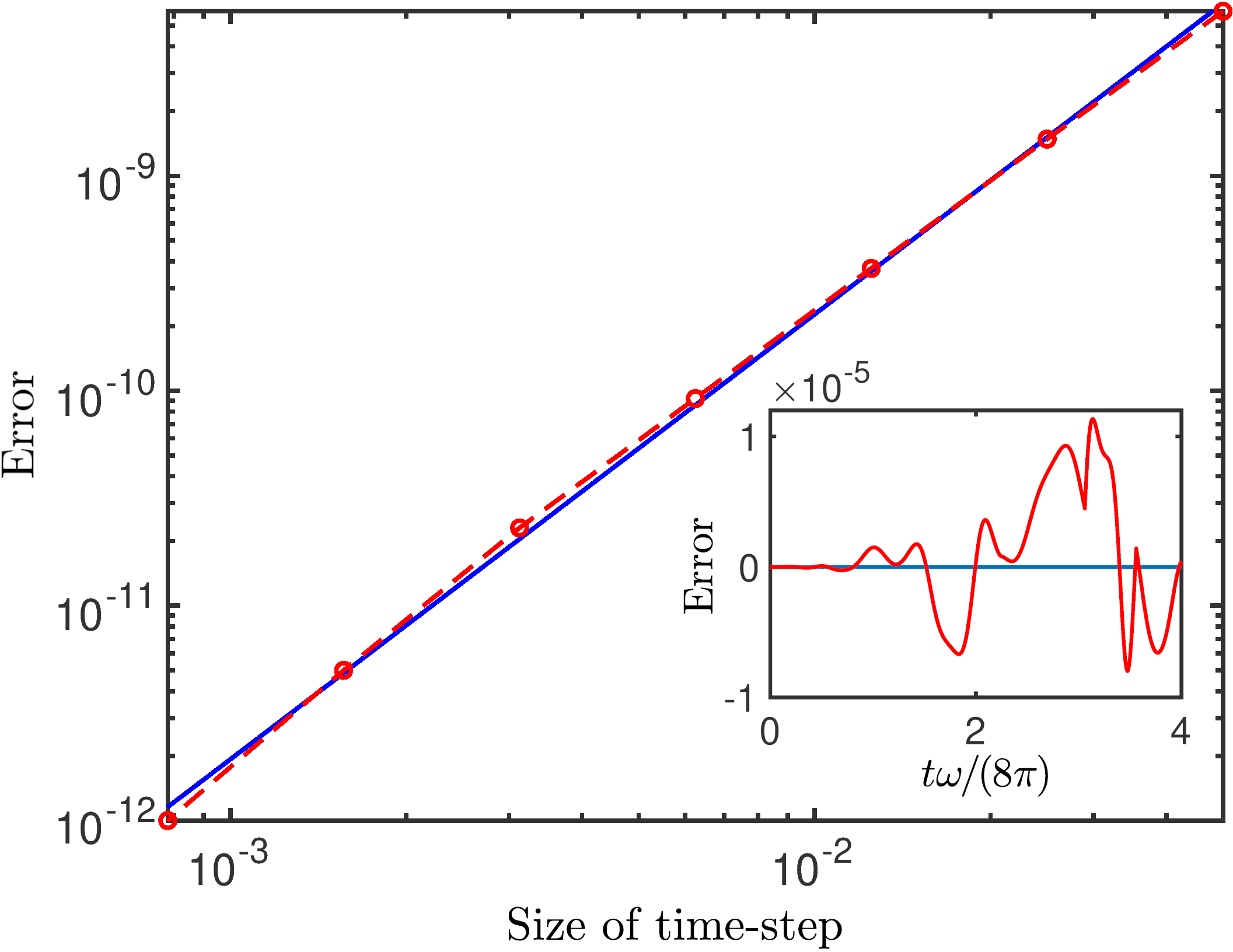}
\caption{(Color online) The error of the numerical solver as a function of the length of the time-step. The line shown in blue for reference has a slope of two, showing that the method is second-order in time. \textit{Inset:} Here, for the control signal shown in Fig.\ref{fig5}, we plot the difference between the evolution of the excited state predicted by the MPS simulations when using a timestep of $T/32000$ and $T/16000$ where $T = 2\pi/\omega$. 
}
\label{fig1app}
\end{figure}

We test the accuracy of the simulation by the usual method of verifying the stability of the solution when simulation parameters are changed. First we use this to check that keeping only the three lowest Fock states for each oscillator is sufficient. We then perform simulations in which we successively halve the time-step and examine the change in the solution over a few time-steps. The results of this procedure, which confirms that the method is second-order in time, are shown in Fig.~\ref{fig1app}.  

Having verified that the numerical method is second-order, we now consider the numerical error of the simulations. Let us denote the units of time used in our simulations by $T$ (this unit is $(2\pi/\omega)$ in which $\omega$ is the transition frequency). For all the simulations presented in the main text we used a time-step of $T/(32000)$. To examine the error we evolve the two-level system under the control signal depicted in Fig.~\ref{fig5} for a time-step of $T/(32000)$ and compare this with the evolution generated by using a time-step of $T/(16000)$. We plot the difference between the populations of the excited state of the qubit for the two evolutions in the inset of Fig.\ref{fig1app}. The error is on the order of $10^{-5}$. For the majority of our simulations we use a damping rate of $10^{-2} (1/T)$, and with the result that the effect of the damping on the system is at the 1\% level. Our simulations therefore report the effect of the bath on the system with an accuracy of approximately $10^{-3}$. This means that we are sensitive to differences between the true evolution and that of the master equations (the ``relative error") to 3 significant figures. Thus when we examine a relative error that is at the 1\% level, this result will itself be accurate to about $\pm 10\%$. When the error between the master equation and the full bath simulations drops below $0.1\%$, we can expect the results will start to be dominated by numerical error.  

The accuracy with which the true joint evolution of the system and bath matches the master equation is limited not only by the finite ratio of the transition frequency to the Rabi frequency, namely $\omega/\Omega$, but also by the ratio of the cut-off frequency to the transition frequency, namely $\Omega_{\ms{cut}}/\omega$, and that of the transition frequency to the damping rate. In Figs.~\ref{fig3} and \ref{fig4} of the main text we see that at small values of the Rabi frequency the relative error stops reducing, reaching a minimum of about $0.1\%$. This error floor is not due to the fact that the system is driven (it is not due to the Rabi frequency since it remains as the driving is reduced) but likely due to either the finite values of $\Omega_{\ms{cut}}/\omega$ and $\omega/\gamma$ and/or the numerical error discussed above. 

\begin{figure}[t] 
\centering 
\leavevmode\includegraphics[width=1\hsize]{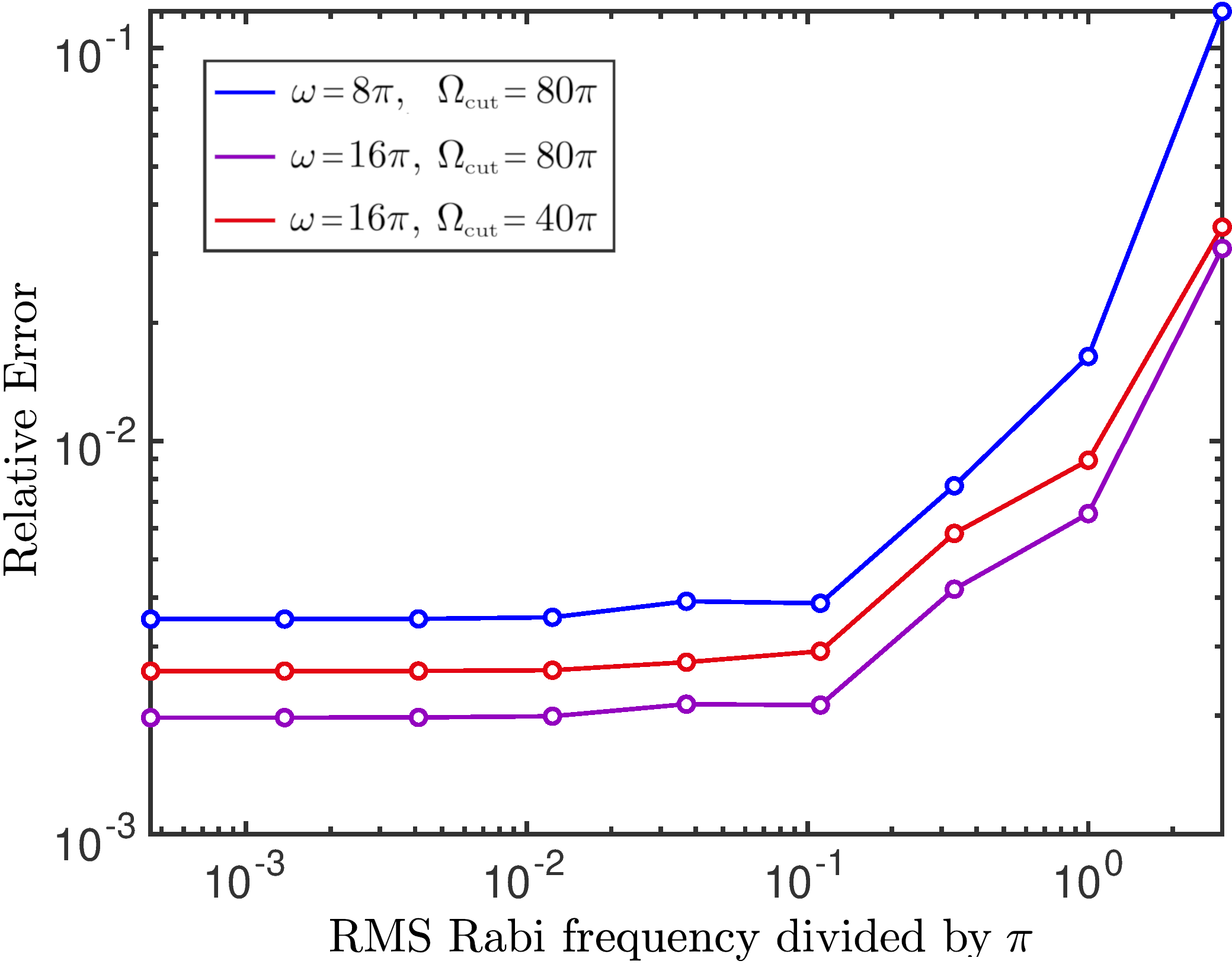}
\caption{(Color online) Here we show three plots of the relative error as a function of the Rabi frequency. The various plots differ by the values of the transition and cut-off frequencies, and show how the floor of the relative error is affected by the finite ratios of $\omega/\gamma$ and $\Omega_{\ms{cut}}/\omega$ (see text). 
}
\label{fig3app}
\end{figure}

In Fig.\ref{fig3app} we explore the origin of the error floor. The blue curve is the relative error as a function of the Rabi frequency with the values for $\omega$ and $\Omega_{\ms{cut}}$ used for the simulations presented in the main text. For the purple curve we doubled the Rabi frequency, so that the resulting drop in the floor indicates that the floor is due in part to the ratio of the transition frequency to the damping rate. For the red curve we both doubled the Rabi frequency and halved the cut-off frequency. The increase in the error floor shows that the finite value of cut-off frequency is also effecting the floor, although less than the damping rate. When we instead keep the Rabi frequency and cut-off the same but reduce the damping rate, we find that the error floor remains essentially unchanged. This indicates that the numerical error is also placing a limit on the floor as we would expect from the discussion above. Note that as $\gamma$ is reduced the effect of the bath on the system is also reduced, thus increasing the effect of the numerical error on the relative error. That is why we see the limiting effect of the numerical error when we reduce $\gamma$, but not immediately when we reduce the transition frequency. Thus the floor in the relative error at approximately $2\times 10^{-3}$ has contributions from the two ratios above as well as the numerical error. 

\section{Numerical methods}
\label{nummeth}

\begin{figure}[t] 
\centering
\leavevmode\includegraphics[width=1\hsize]{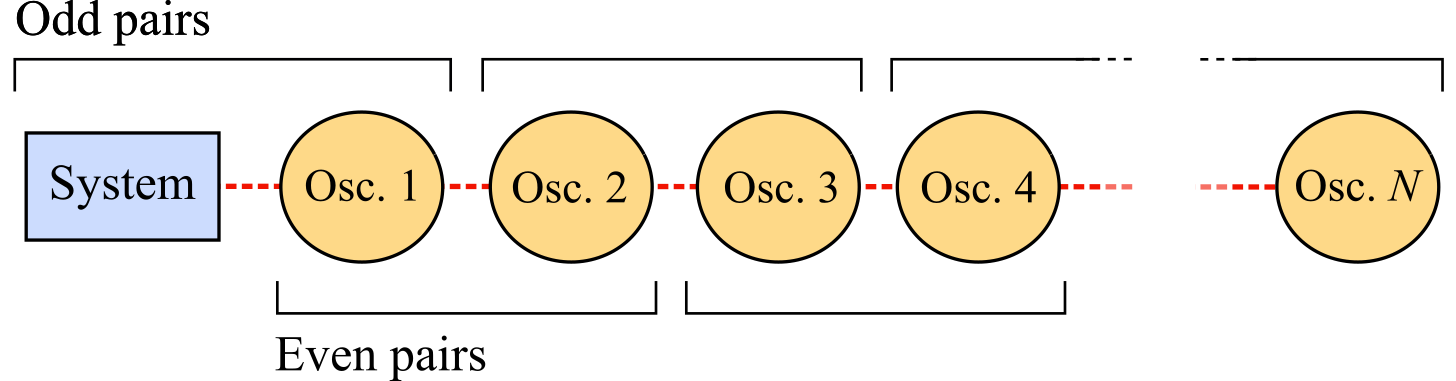}
\caption{(Color online) A depiction of the chain configuration whose dynamics is equivalent to that of a system coupled to a bath of independent oscillators.
}
\label{fig4app}
\end{figure}

To simulate two- and three-level systems coupled to a bath of Harmonic oscillators we use the method first developed by Bulla \textit{et al.\ }\cite{Bulla03, Anders07, Bulla08} in which the system and bath are mapped to a semi-infinite chain of oscillators in which the system is coupled only to the first oscillator (see Fig.~\ref{fig4app}). We use a refinement of this method developed by Chin \textit{et al.\ }\cite{Chin10,Prior10} which exploits an exact analytic mapping between the bath and the oscillator chain. The system and chain of oscillators is then simulated using the matrix-product-state (MPS) method developed by Vidal~\cite{Vidal03, Vidal04}. 

To efficiently evolve a nearest-neighbor chain of systems stored as a matrix-product state one must evolve nearest-neighbor pairs separately~\cite{Vidal03}. Simulating the dynamics of the chain thus requires splitting the evolution in each time-step into at least two parts, where the first (part A) evolves all ``odd" pairs and the second (part B) all ``even" pairs (see Fig.~\ref{fig4app}). To obtain a method that is more than first-order in time requires splitting each time-step into more than one application of $A$ and $B$. Our analysis requires high accuracy because the damping induced in the system is small to begin with, and it is small errors in the ability of the master equation to simulate the effect of this damping that we wish to quantify. We therefore require to go beyond a first-order method, and because our Hamiltonian is time-dependent we cannot use the split-operator methods commonly employed in MPS simulations~\cite{Suzuki85, De_Leo2016}. Here we take advantage of the fact that only the system (and thus part A) is time-dependent, and use the following fairly simple split-operator construction that realizes Heun's 2nd-order Runge-Kutta method~\cite{Mayers03}. 

The equation we wish to evolve has the form 
\begin{align}
    \dot{x}= [A(t)+B]x 
\end{align}
in which $A(t)$ does not commute with $B$ or with itself at different times. Defining 
\begin{align}
    A_1 & \equiv A(t) \\
    A_2 & \equiv A(t+\Delta)
\end{align}
we evolve $x(t)$ using three operators: 
\begin{align}
    x(t + \Delta t) & = \left[ 1+\frac{\Delta A_2}{2} + \frac{\Delta^2 A_2 A_1}{8} \right] \nn \\ 
    & \;\;\;\;\; \times e^{\Delta B} \left[ 1+\frac{\Delta A_1}{2} + \frac{\Delta^2 A_2 A_1}{8} \right] . 
\end{align}

More elaborate split-operator methods for situations in which only $A$ is time-dependent can be found in~\cite{Bandrauk92}. The chain topology in which the evolution involves pairs of oscillators lends itself to parallelization, in which the chain is divided among a large number of processors. This parallelization allows us to readily simulate chains with thousands of oscillators.

\end{document}